\documentstyle{article}
\def\PL #1 #2 #3 {{\rm Phys. Lett.} {\bf#1} (#3) #2}
\def\NP #1 #2 #3 {{\rm Nucl. Phys.} {\bf#1} (#3) #2}
\def\ZP #1 #2 #3 {{\rm Z. Phys.} {\bf#1} (#3) #2}
\def\PRL #1 #2 #3 {{\rm Phys. Rev. Lett.} {\bf #1} (#3) #2}
\def\PR #1 #2 #3 {{\rm Phys. Rev.} {\bf#1} (#3) #2}
\def\MPL #1 #2 #3 {{\rm Mod. Phys. Lett.} {\bf#1} (#3) #2}
\def\RMP #1 #2 #3 {{\rm Rev.~Mod. Phys.} {\bf#1} (#3) #2}
\def\ifm{\ifmmode}

\newcommand{\notp}{\ \hbox{{$p$}\kern-.43em\hbox{/}}}
\newcommand{\notE}{\ \hbox{{$E$}\kern-.43em\hbox{/}}}
\newcommand{\beq}{\begin{equation}}
\newcommand{\eeq}{\end{equation}}
\newcommand{\beqn}{\begin{eqnarray}}
\newcommand{\eeqn}{\end{eqnarray}}
\newcommand{\beqs}{\begin{eqnarray*}}
\newcommand{\eeqs}{\end{eqnarray*}}

\def\ts2p3{s_{2'3}}

\def\be{\begin{equation}}
\def\ee{\end{equation}}
\def\bea{\begin{eqnarray}}
\def\eea{\end{eqnarray}}

\begin{document}
\thispagestyle{empty}
\begin{flushright}
\hfill{FERMILAB-CONF-96/307-T}
\end{flushright}

\vskip 2cm
\begin{center}
$M_W$ MEASUREMENT AT THE TEVATRON WITH HIGH LUMINOSITY~\footnote{
Talk given by S. K. at the DPF96 Conference, Minneapolis, 
MN, August~10--15, 1996, to appear in the Proceedings
 }
\vglue 1.4cm
\begin{sc}
Walter T. Giele and Stephane Keller\\
\vglue 0.2cm
\end{sc}
{\it Fermilab, MS 106\\
Batavia, IL 60510, USA}
\vglue 0.5cm
\end{center}
\vglue 1.5cm
\begin{abstract}
\par \vskip .1in \noindent
We present an alternate method to measure $M_W$ at the Tevatron with 
high luminosity from a direct comparison of the
$W$ and $Z$ distributions.  
\end{abstract}
\newpage

Currently, the mass of the $W$-boson ($M_W$) is measured at the Tevatron from
the transverse mass distribution ($M_T$).
At higher luminosity, 
the uncertainty is expected to scale 
with the inverse of the square root of the integrated luminosity, 
as most of the systematic uncertainties are controlled by data samples.  
However, a recent study~\cite{TEV2000} has shown that  
the increase in the number of interactions per crossing ($I_C$) will 
substantially degrade the uncertainty in the reconstruction of the 
transverse energy of the neutrino($P_{T\nu}$), 
and therefore the uncertainty with which $M_W$ can be extracted.  
In Table 1, we reproduce 
the expected uncertainty calculated in that 
study at $1fb^{-1}$($I_C=3$) and $10fb^{-1}$($I_C=9$).  Note that no detector
upgrades were considered for that analysis.
\begin{table}[b]
\begin{center}
\begin{tabular}{|c|c|c|} \hline 
  $\Delta M_W$           & $\int L dt= 1 fb^{-1}$, $I_C=3$    
& $\int L dt=10fb^{-1}$, $I_C=9$    \\ \hline
  statistical        & 29      & 17\\ \hline
  systematic   & 42      & 23\\ \hline
  total        & 51    &29\\ \hline
Benchmark & 25 & 7.9 \\ \hline
\end{tabular}
\end{center}
\caption{\it
Projected uncertainty on $M_W$, along with our benchmark, 
in $MeV/c^2$, per experiment.  The $e$ and $\mu$ channels are both included.    
}
\label{tab:mw}
\end{table}
For comparison, we use as a benchmark the current 
CDF uncertainty of $180MeV/c^2$ at about $20 pb^{-1}$ simply scaled 
with the luminosity, see Table 1.  
The uncertainties at $1fb^{-1}$ and $10fb^{-1}$ are about 2 and 4 times worse 
than our benchmark, and are dominated by the systematic uncertainty.    
It would be interesting to have a method that is dominated by 
statistical uncertainty.
The total uncertainty should be compared to the expected 
uncertainty at LEPII of $40 MeV/c^2$ for the 4 detectors combined. 
Note also that the TeV33 committee report~\cite{TEV33} suggests a target 
of $30fb^{-1}$ by the end of 2006, with a goal of $\Delta M_W = 15MeV/c^2$.  
Recently~\cite{KW96}, the prospect to measure $M_W$ at the LHC was 
investigated, and no problems that would prevent a very precise 
determination were uncovered. 

One possible solution to the multiple interactions per crossing problem 
is to divide the data sample into subsamples corresponding to fixed $I_C$ 
and to study the effect~\cite{CAR}.  Another solution would be to lower 
the bunch spacing, in order to reduce $I_C$.
However, this would require detector upgrades beyond what is currently 
planned.  Finally, observables that do not depend on $P_{T\nu}$, like
the momentum ($P_l$) or the transverse 
momentum ($P_{Tl}$) of the charged lepton, could be used.  
In this short contribution we concentrate on this latter solution.  

First, let us consider $P_l$.  
A few years ago, a study was performed during the Madison-Argonne
workshop~\cite{PER}, and it was concluded that the total uncertainty 
using this observable is about 1.5 times worse than our benchmark. 
It is reasonable to assume that for this observable the uncertainty 
will scale normally to higher luminosity, such that this method
could provide a better measurement than using the $M_T$ distribution.  
This analysis should be repeated as it is not clear if all the uncertainties
were accounted for.  

\begin{figure}[t]
\vglue 4.3cm
\vbox{\includegraphics{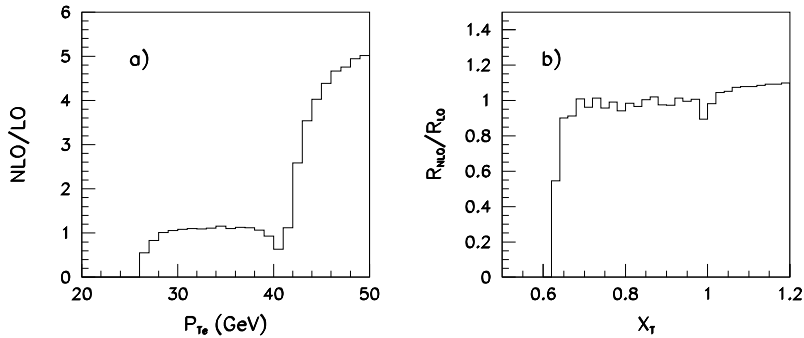} }
\caption{\it
$W$ production.  a) Ratio of NLO over LO cross sections 
as a function of $P_{Te}$.  
b) Ratio of NLO over LO $R$ as a function of $X_T$.}   
\label{fig:fig1}
\end{figure}

Let us now turn to $P_{Tl}$.  In Fig.~\ref{fig:fig1}a, 
we present for $W$ production the ratio of the QCD next-to-leading order (NLO) 
calculation over the leading order (LO) calculation as a function of 
$P_{Te}$.  
As can be seen, there are large corrections in the region 
of interest, around $M_W/2$.  As a result, the perturbative 
expansion can not be trusted, these large corrections need to be resummed.   
Here, we want to suggest an alternative to resummation by   
considering the ratio of $W$ over $Z$ distributions.  The basic 
idea is that the large corrections are universal and cancel in the ratio.    
The $P_{Tl}$-distributions of the $W$ and $Z$ peak at different places,
at about half the vector boson mass($M_V$), 
such that the first step is to consider scaled variables:
$X_T = P_{Tl}/(M_V/2)$.  The $X_T$-distributions have also 
large QCD corrections.  
We define $R$ as the ratio of the 
$W$ over $Z$ $X_T$-distributions.
The ratio of NLO over LO of $R$ is presented in Fig.~\ref{fig:fig1}~b.  
As can be seen, the corrections are small and of the order of
10-20 \% which indicates that the perturbative expansion for this
observable is well behaved.
The mass dependence mainly enters when the $P_{Tl}$ distribution 
is transformed into the $X_T$ distribution, and $M_W$ can be measured 
by fitting the ratio $R$.

The limitation of the method
is that it depends on the $Z$ statistics which is about 
$5$ times smaller than the $W$ (considering that both the electron 
and the positron can be used in the $Z$ case).
Therefore, a statistical uncertainty about $2$($\sim \sqrt{5}$) times worse 
than our benchmark can be expected.  
Note that the method still
depends on the $P_{T\nu}$ for the identification of the W, but  
this dependence can be reduced by imposing a cut on $P_{Tl}$ bigger 
than on $P_{T\nu}$.

There are several advantages to the method.  First, it only uses the NLO QCD 
calculation (the NNLO could be used if it becomes available), 
there is no need for any resummation.  
Second, $M_W$ is directly measured with respect to $M_Z$ which has been
measured very precisely at LEP.  Finally, 
because of the use of the ratio, the systematic uncertainty
is expected to be small.  Only the systematic effects that are different 
for the $W$ and $Z$ should contribute, like the isolation criteria of the 2nd
electron in the $Z$ case, or some of the backgrounds.
Considering a small systematic uncertainty, overall the ratio 
method should give an uncertainty on $M_W$ smaller than 
$2$ times worse than our benchmark.  Therefore, it has the potential to 
do better than the conventional transverse mass method.  Finally note that 
the ratio method can be used with any other observables, like $M_T$ itself or
$P_l$.

In conclusion, at this point there is no clear winner at high 
luminosity between the different observables to measure $M_W$, 
and 
the direct comparison of $W$ and $Z$ distribution 
seems very promising. 

\end{document}